\documentclass[lettersize,journal]{IEEEtran}
\usepackage{amsmath,amsfonts}
\usepackage{algorithmic}
\usepackage{algorithm}
\usepackage{array}
\usepackage{textcomp}
\usepackage{stfloats}
\usepackage{url}
\usepackage{verbatim}
\usepackage{graphicx}
\usepackage{cite}
\usepackage{multirow}
\usepackage{subcaption}
\begin{document}

\title{Investigation of Japanese PnG~BERT language model in  text-to-speech synthesis for pitch accent language}

\author{Yusuke Yasuda,~\IEEEmembership{Member,~IEEE,} and Tomoki Toda~\IEEEmembership{Member,~IEEE} 
\thanks{This paper was produced by the IEEE Publication Technology Group. They are in Piscataway, NJ.}
\thanks{Manuscript received April 19, 2021; revised August 16, 2021.}}

\markboth{Journal of \LaTeX\ Class Files,~Vol.~14, No.~8, August~2021}%
{Shell \MakeLowercase{\textit{et al.}}: A Sample Article Using IEEEtran.cls for IEEE Journals}

\IEEEpubid{0000--0000/00\$00.00~\copyright~2021 IEEE}

\maketitle

\begin{abstract}
End-to-end text-to-speech synthesis (TTS) can generate highly natural synthetic speech from raw text. However, rendering the correct pitch accents is still a challenging problem for end-to-end TTS. To tackle the challenge of rendering correct pitch accent in Japanese end-to-end TTS, we adopt PnG~BERT, a self-supervised pretrained model in the character and phoneme domain for TTS. We investigate the effects of features captured by PnG~BERT on Japanese TTS by modifying the fine-tuning condition to determine the conditions helpful inferring pitch accents. We manipulate content of PnG~BERT features from being text-oriented to speech-oriented by changing the number of fine-tuned layers during TTS. In addition, we teach PnG~BERT pitch accent information by fine-tuning with tone prediction as an additional downstream task. Our experimental results show that the features of PnG~BERT captured by pretraining contain information helpful inferring pitch accent, and PnG~BERT outperforms baseline Tacotron on accent correctness in a listening test.
\end{abstract}


\section{Introduction}
\IEEEPARstart{E}{nd}-to-end (E2E) text-to-speech synthesis (TTS) can generate highly natural synthetic speech from raw texts \cite{Shen2017, DBLP:journals/corr/abs-1809-08895}. However, rendering correct pitch accents or tones, which are accents involving pitch change, remains a challenging problem for E2E-TTS \cite{DBLP:conf/icassp/YasudaWTY19, Li2019, fujimoto19_ssw}. The accuracy of rendering pitch accents is crucial in pitch accent languages such as Japanese, because pitch accents control the meaning of words. Conventional TTS systems resolve accent information with a morphological analyzer-based text front-end by looking it up in a accent dictionary, which is normally expensive and requires language-specific knowledge to construct \cite{HTSWorkingGroup2014JOPEN}. 
The characteristics of E2E-TTS to use texts or phonemes directly as input enables it to be applied to various data and languages without constructing a pronunciation dictionary.
On the other hand, the challenge of rendering a pitch accent in an E2E-TTS approach comes from building knowledge corresponding to accent dictionary implicitly from text and speech pair.

To tackle the challenge of rendering the correct pitch accent in Japanese E2E-TTS, two main issues should be addressed. One is the low word coverage in the speech corpus. Pitch accents in pitch accent languages are mainly determined by words, so a low word coverage in training speech data can result in the main problem of poor pitch accents for an end-to-end TTS model. However, even a large-scale speech corpus can never match a lexicon dictionary in terms of word coverage \cite{Taylor2019}. The other problem is the diversity of characters. The Japanese writing system contains ideographic characters that represent meaning rather than pronunciation. Owing to the diversity of ideographic characters, Japanese end-to-end TTS from raw texts has been impossible. Therefore, phonemes are used as input in end-to-end Japanese TTS instead. This results in loss of word-related information, leading to incorrect pitch accents \cite{DBLP:conf/icassp/YasudaWTY19, fujimoto19_ssw}.
Both phonemes and pitch accents are important speech representations to render in Japanese TTS, but these two representations depend on different features: rendering phonemes depends on characters and phonemes, and rendering pitch accents depends on word and syntactic features. Therefore, a framework to capture both surface form and high-level features is required.

To tackle these two main issues, we utilize a self-supervised pre-trained model using a large-scale text corpus. As a self-supervised learning method, we use PnG~BERT \cite{jia21_interspeech}. PnG~BERT is an extension of BERT \cite{DBLP:conf/naacl/DevlinCLT19} designed for TTS as a downstream task by capturing both word and phoneme contexts. We adopt PnG~BERT for Japanese end-to-end TTS because (1) it can capture both word and phoneme contexts; (2) word and phoneme alignment can be learned in text domain instead of the speech domain; (3) a downstream task can be performed against features corresponding to a phoneme segment instead of a word segment. Regarding (1), we can expect that features extracted from PnG~BERT are helpful to render correct pitch accents, because Japanese pitch accents are mainly determined by words. Moreover, BERT can not only provide semantic information but also syntactic information \cite{DBLP:conf/acl/JawaharSS19, DBLP:conf/acl/TenneyDP19}.  Syntactic information such as conjugation type is also helpful inferring the fundamental frequency ($F_o$) in Japanese speech \cite{wangPhD}. Concerning (2), introducing multiple soft-attention layers to TTS enables to us align multiple linguistic units to speech \cite{Aso2020} and to overcome the lack of word boundary symbols in the Japanese writing system. However, it is preferable to align multiple language units in the text domain considering the limited volume of clean speech data. On the other hand, PnG~BERT can learn alignment between words and phonemes in self-attention layers \cite{Vaswani2017} by utilizing large-scale text corpus \cite{jia21_interspeech}. This characteristic enables feature extraction considering both words and phonemes for TTS. With (3), we can introduce tone prediction as a downstream task of PnG~BERT in addition to TTS straightforwardly. Tone representation of Japanese pitch accent is determined in at the syllable level. Therefore the tone can be predicted in the phoneme part of PnG~BERT token-by-token. Naturally, we can expect that tone prediction is also benefited from word or syntactic information captured by PnG~BERT considering (1). The tone prediction is used to explicitly teach PnG~BERT accent information in this work.

In this study, we investigate the effects of feature contents captured by PnG~BERT on Japanese TTS by modifying the fine-tuning condition. 
In our experiments, we manipulate the content of PnG~BERT features from being text-oriented to speech-oriented by changing the number of fine-tuned layers by TTS. In addition, we inject pitch accent information itself to PnG~BERT features by fine-tuning with tone prediction as an additional downstream task.

This paper is organized as follows. In Section \ref{sec:background}, we describe backgrounds of BERT, PnG~BERT and background of Japanese TTS. In Section \ref{sec:japanese-tts-pngbert} we introduce our proposed method for Japanese TTS using PnG~BERT. Section \ref{sec:experiments} shows our experimental results. In Section \ref{sec:related-works}, we summarize related works. Finally, In Section \ref{sec:conclusion}, we conclude our findings.

\section{Background}
\label{sec:background}

\subsection{BERT}

BERT \cite{DBLP:conf/naacl/DevlinCLT19} is a self-supervised learning-based language model for general language representation. It has two steps to learn textual representation and task-related features: pre-training and fine-tuning. BERT uses the masked language model (MLM) objective during pre-training to capture general language representation. The MLM objective optimizes a model by predicting masked tokens given input texts in which some parts of the tokens are randomly masked. In addition to MLM, BERT has the next sentence prediction (NSP) objective, which classifies whether two sentences are adjacent or unrelated. During fine-tuning, BERT uses a supervised objective to solve a specific task. Examples of a downstream task to fine-tune BERT are natural language processing (NLP) tasks such as language understanding and question answering.

To use the MLM and NSP objectives, BERT arranges input tokens as follows: an input sequence always starts with a \texttt{CLS} token, which is used for a classification task such as NSP. Each input sentence ends with a \texttt{SEP} token, which means the end of a sentence. As for the masking method, a portion (e.g., 15\%) of the input tokens are randomly chosen to be predicted as targets in self-supervised learning. Most of the selected tokens (e.g., 80\%) are masked by being replaced with a \texttt{MASK} token Some tokens (e.g., 10\%) are replaced with a random token, whereas some tokens (e.g., 10\%) are not replaced. A BERT model is trained to recover the original sentences from the modified inputs.

The architecture of BERT is equivalent to that of the encoder of Transformer \cite{Vaswani2017}. It consists of a stack of self-attention blocks that are a combination of a multihead self-attention layer and a feed-forward network with ReLU activation. Positional encoding is used to take the order of tokens into consideration.

\subsection{Png BERT}

Language models for TTS tasks aim to capture pronunciation-related information, whereas language models for NLP tasks focus on syntactic or semantic information. PnG~BERT is a self-supervised learning-based language model that can capture both contextual word and phoneme information in the text domain \cite{jia21_interspeech}. PnG~BERT is an extended method of BERT: it uses the same Transformer-based architecture and MLM training objective. PnG~BERT extends BERT in terms of input representation, masking strategy, and fine-tuning to capture pronunciation information.

\begin{figure}[!t]
    \begin{center}
    \begin{subfigure}[t]{\columnwidth}
    {\includegraphics[width=\columnwidth]{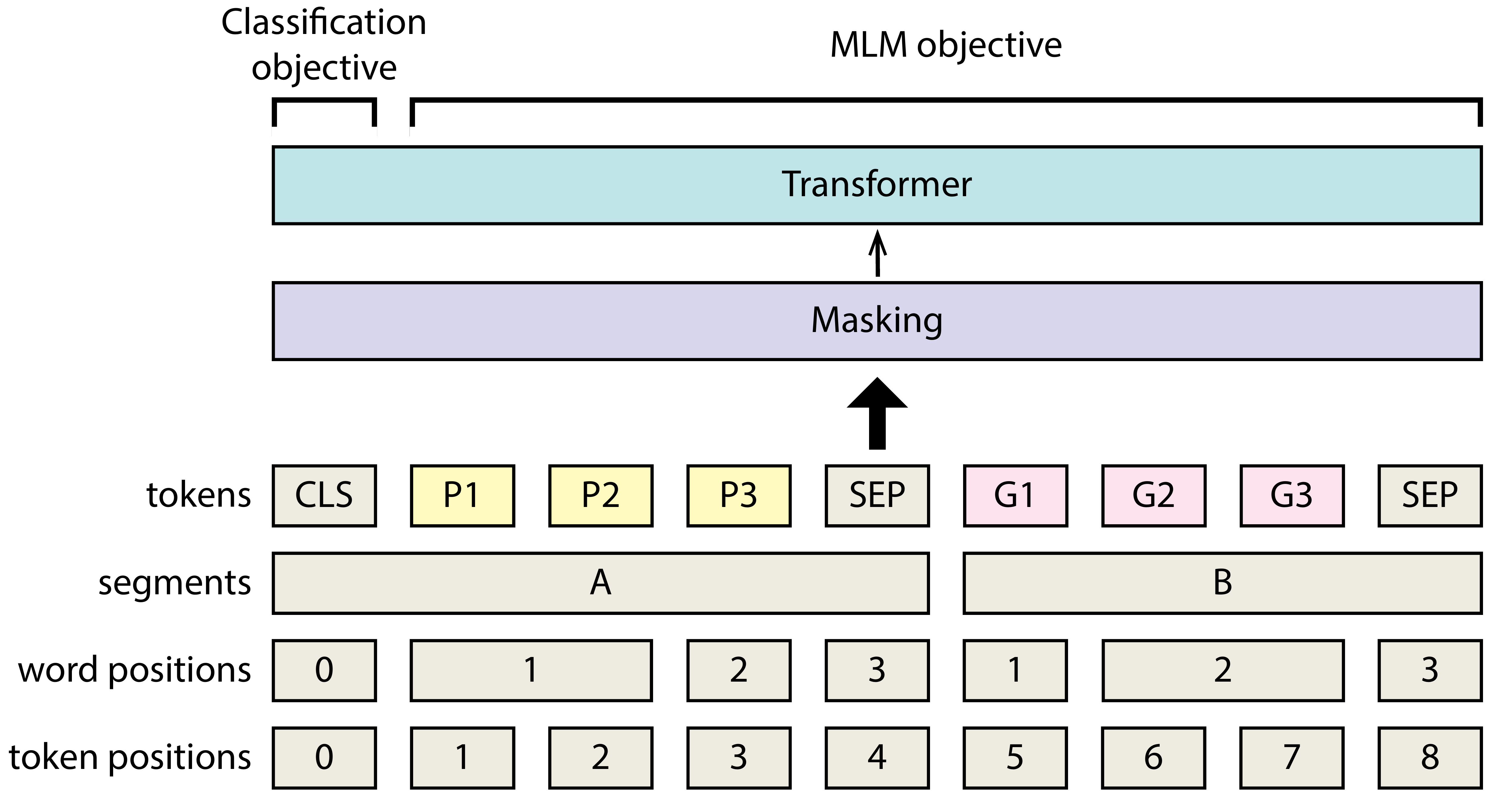}}
    \subcaption{Pre-training}
    \end{subfigure}
    \vspace{2mm}
    \begin{subfigure}[t]{\columnwidth}
    {\includegraphics[width=\columnwidth]{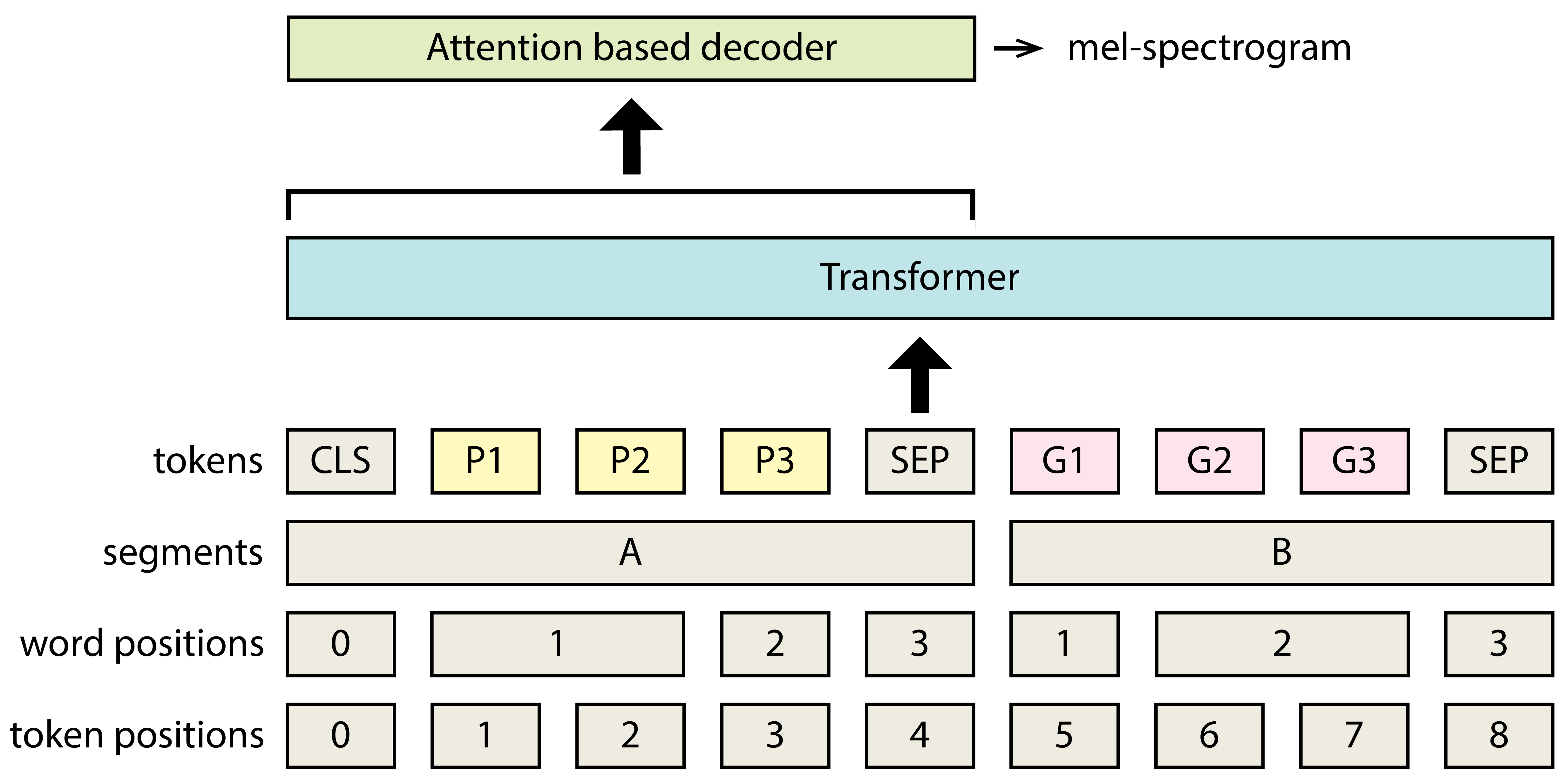}}
    \subcaption{Fine-tuning}
    \end{subfigure}
    \end{center}
    \caption{Mechanism of PnG~BERT. P: phoneme, G: grapheme.}
    \label{fig:pngbert-trainig}
\end{figure}

Figure \ref{fig:pngbert-trainig} shows the structure of PnG~BERT. It uses graphemes and phonemes as input representation instead of words. Here, we use graphemes to refer to as any textual representation unit such as characters and subwords. A target downstream task of PnG~BERT is TTS. Character and phonemes are common input representations for end-to-end TTS. The input sequence for PnG~BERT consists of the concatenation of phoneme and grapheme sequences of a corresponding text. Each of the phoneme and grapheme sequences ends with a \texttt{SEP} token. A \texttt{CLS} symbol is prepended to the whole sequence. The \texttt{CLS} token is for the sentence classification task. Token, word, and segment positions are specified by positional encoding \cite{Vaswani2017}, because self-attention in Transformer is permutation invariant.

The masking strategy of PnG~BERT is designed at the word level. Because the input representations in PnG~BERT are graphemes and phonemes, masking at token-level limits contextual information captured in pretraining up to surface-form level. The word-level masking enables PnG~BERT  to learn all layers of contextual information: semantic, surface form, and pronunciation. In PnG~BERT, one of the following masking strategies is chosen for a random word: both graphemes and phonemes are masked; only graphemes are masked; only phonemes are masked; both graphemes and phonemes are kept intact; both graphemes and phonemes are randomly replaced with other phonemes and graphemes. For evaluation, grapheme-to-phoneme (G2P) and phoneme-to-grapheme (P2G) masking strategies may be used in addition to MLM. In the G2P masking, all the phoneme segments are masked. In P2G masking, all the grapheme segments are masked.

A TTS downstream task can be conducted with PnG~BERT by feeding its final outputs corresponding to the phoneme segment into the TTS decoder. PnG~BERT models are jointly fine-tuned with the TTS decoder while freezing its parameters in part of its layers to prevent the loss of learned knowledge during pre-training. PnG~BERT is compatible with any TTS decoder.

\subsection{Japanese TTS}

Japanese is classified as a pitch accent language in terms of prosodic description. Japanese pitch accents are lexical, which means that accents depend on words or a combination of words to form an accentual phrase. The Japanese pitch accent has an accentual nucleus position that is specified by a pitch fall within an accent phrase. The accent nucleus position is measured in a syllable unit called mora. Although the pitch accent of individual words can be described in a lexicon, it is affected by adjacent words in an accentual phrase, which is a phenomenon called accent sandhi.

The Japanese TTS system uses text front-end for word segmentation and lexicon lookup for pronunciation by using a morphological analyzer. After the lexicon lookup, accent sandhi is estimated on the basis of rules \cite{Sagisaka1983} or machine learning \cite{MasayukiSUZUKI20172016AWI0004, Minematsu2007}. The Japanese text front-end provides full-context labels that include various type of information helpful to TTS such as part of speech. Ref. \cite{YASUDA2021101183} shows that at least accent nucleus position, or the accentual type, should be provided to TTS to render a natural pitch accent.

\section{Japanese TTS using Png BERT}
\label{sec:japanese-tts-pngbert}

\begin{figure}[!t]
    \begin{center}
    \begin{subfigure}[t]{\columnwidth}
    {\includegraphics[width=\columnwidth]{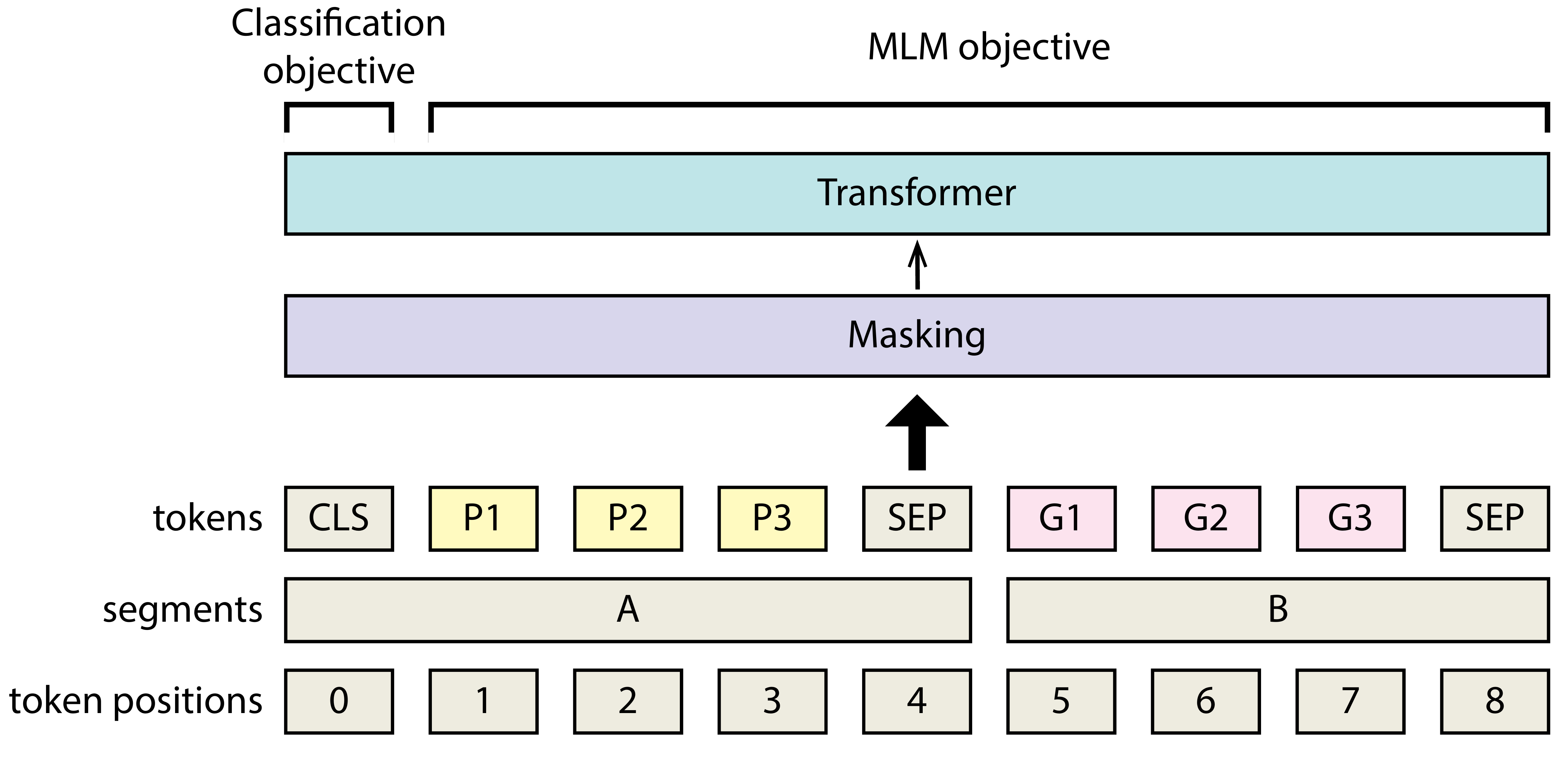}}
    \subcaption{Pre-training}
    \end{subfigure}
    \vspace{2mm}
    \begin{subfigure}[t]{\columnwidth}
    {\includegraphics[width=\columnwidth]{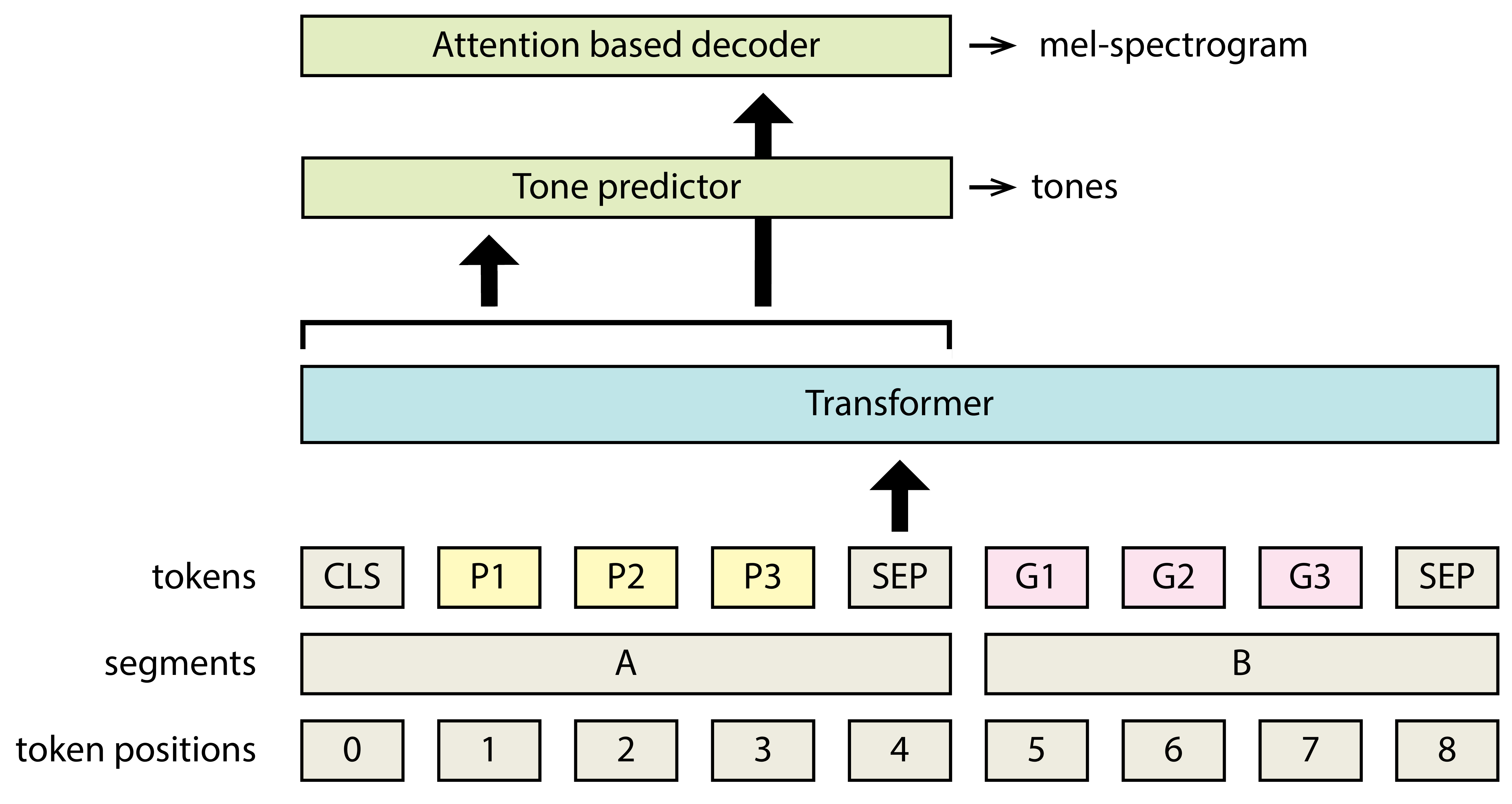}}
    \subcaption{Fine-tuning}
    \end{subfigure}
    \end{center}
    \caption{Mechanism of Japanese PnG~BERT to perform pre-training and fine-tuning with text-to-speech synthesis and tone prediction as downstream task. P: phoneme, G: grapheme.}
    \label{fig:pngbert-japanese}
\end{figure}

Figure \ref{fig:pngbert-japanese} shows our method to conduct Japanese TTS using PnG~BERT. Its general framework is the same as that of the plain PnG~BERT: graphemes and phonemes are concatenated as one sequence, and the sequence is fed to Transformer layers; the PnG~BERT model is pretrained with the MLM objective by masking a grapheme and phoneme pair at the word level; during fine-tuning and speech generation, the phoneme part of the output of the last Transformer layer is used as the input for a TTS decoder to generate a spectrogram. 

We introduce two changes to the PnG~BERT to conduct Japanese TTS. The first modification is the absence of word-level alignment information between graphemes and phonemes as word positions in the inputs. The Japanese writing system lacks clear word boundary symbols. To obtain word boundaries of Japanese texts used for pre-training, we use morphological analyzers. Obviously, automatically derived word boundaries from the morphological analyzer contain errors. We do not use word positions to extract features to avoid dependence on erroneous information. Thus, word-level alignment information is required only during pre-training to conduct word-level masking. We expect that removing word positions does not affect the performance of TTS negatively as Ref. \cite{jia21_interspeech} suggests. We also expect that masking based on erroneous word boundaries derived from the morphological analyzer does not have major negative effects, because it just affects size of word units but does not affect consistency of word alignments between graphemes and phonemes. It is known that the consistency of the word-level masking affects the performance of pre-training and the downstream TTS task in PnG~BERT \cite{jia21_interspeech}. We will confirm these assumptions by measuring the G2P conversion performance of PnG~BERT.

The second modification is the introduction of tone prediction to TTS as an additional downstream task. Japanese pitch accents are an important spoken feature to disambiguate homonyms. Unfortunately, accent labels are not available on the scale of large corpora that are used for pre-training. Therefore, we incorporate tone prediction as one of the downstream tasks. The tone prediction task is performed against the phoneme part of PnG~BERT's output features similar to the TTS downstream task. We use tone labels to represent accent information because they are aligned to phonemes in a syllable unit. Ideally, the relationship between extracted features and pitch accents is to be learned implicitly from the target mel-spectrogram that contains $F_o$ information. The tone prediction task is to explicitly teach PnG~BERT pitch accents. We expect that fine-tuning PnG~BERT with accent prediction makes it easy for PnG~BERT to learn pitch accents.

\section{Experimental Evaluations}
\label{sec:experiments}

To investigate the effects of feature contents captured by PnG~BERT on Japanese TTS, we pre-trained a Japanese PnG~BERT model and constructed TTS models under various fine-tuning conditions.

\subsection{Pre-training condition}

To construct a PnG~BERT model, we used a BERT model with model size, so-called BERT-base \cite{DBLP:conf/naacl/DevlinCLT19}, which consists of 12 Transformer layers with 768 hidden size \footnote{We also tried the original model size, which consists of 6 Transformer layers with 512 hidden size. We found that fine-tuning of PnG~BERT with the original model size converged faster and its synthetic speech sounded similar quality as PnG~BERT with BERT-base size.}. The model is trained up to 11.63M steps with 20 batch size and learning rate with linear decay started from $5.0 \times 10^{-5}$ with the Adam optimizer with L2 regularization \cite{DBLP:journals/corr/KingmaB14}. Note that the original PnG~BERT uses the SM3 optimizer\cite{DBLP:conf/nips/Anil0KS19} instead of Adam. We found that with Adam, it was easy to optimize the model, whereas with SM3, it was difficult to find the proper hyper parameters, although it was memory-efficient.

We randomly selected 25\% words in a sentence as a target of the MLM objective. For the 25\% selected words, we applied the following masking strategy: (a) both graphemes and phonemes are masked for 48\% random words; (b) only the character part is masked for 8\% random words; (c) only the phoneme part is masked for 8\% random words; (d) both characters and phonemes are kept intact for 10\% random words; (d) both characters and phonemes are randomly replaced with other phonemes or characters for 10\% random words. Note that (b) and (c) are not present in the original PnG~BERT \cite{jia21_interspeech}. In preliminary experiment, we found that these masking strategies were very effective for improving G2P and P2G performances.

We followed the input representations of the original PnG~BERT paper \cite{jia21_interspeech}. We used character and phoneme representations as input to PnG~BERT. The two sequences were appended with the \texttt{SEP} token and concatenated into one sequence. The \texttt{CLS} token was prepended, but we did not use the NSP objective.

As Japanese text corpora for pre-training, we used Wikipedia\footnote{https://dumps.wikimedia.org/} and Aozorabunko\footnote{https://github.com/aozorabunko/aozorabunko/}. Table \ref{tbl:text-corpus} shows statistics of the text corpora used in this study. Wikipedia is a collection of online encyclopedia. We used a subset of Japanese Wikipedia that contained about 18.5M sentences. Aozorabunko is collection of Japanese books in public the domain. We used subset of Aozorabunko that contained about 4.9M sentences. Phoneme transcriptions were obtained by morphological analysis using the Kuromoji analyzer\footnote{https://github.com/atilika/kuromoji} with Neologd dictionary \cite{sato2015mecabipadicneologd,sato2016neologdipsjnl229,sato2017mecabipadicneologdnlp2017}. For validation, we used text and phoneme labels from the JSUT corpus \cite{DBLP:journals/corr/abs-1711-00354}, which was also used to fine-tune a TTS model. We used a subset of JSUT corpus called \texttt{basic5000}, which contains 5,000 sentences that cover the pronunciations of commonly used ideographic characters.

\begin{table}[t]
\caption{Statistics of Japanese text corpus used in this study.}
\label{tbl:text-corpus}
\begin{center}
\begin{tabular}{|crcc|}\hline
\multirow{2}{*}{Corpus} & \multirow{2}{*}{\#Records} & \multicolumn{2}{c|}{Usage} \\
~ & ~ & Pre-training & Fine-tuning \\\hline
\texttt{Wikipedia} & 18,486,538 & Training & - \\
\texttt{Aozorabunko} & 4,891,288 & Training & - \\
\texttt{JSUT} & 6,385 & Validation & Training/Validation/Test\\\hline

\end{tabular}
\end{center}
\end{table}

\subsection{TTS Systems}
\label{subsec:tts-systems}

We constructed five TTS systems using the same pre-trained PnG~BERT model on the basis of different fine-tuning conditions. 
\begin{itemize}
    \item[1)] \texttt{PGB0}: No fine-tuning
    \item[2)] \texttt{PGB2}: Two-layer fine-tuning
    \item[3)] \texttt{PGB4}: Four-layer fine-tuning
    \item[4)] \texttt{PGB6}: Six-layer fine-tuning
    \item[5)] \texttt{PGB2T}: Two-layer fine-tuning with tone prediction in addition to TTS
    \item[6)] \texttt{PGB2MC}: Two-layer fine-tuning and all graphemes are masked during fine-tuning and prediction.
\end{itemize}

The \texttt{PGB0} system is equivalent to TTS using fixed features extracted from PnG~BERT. The \texttt{PGB2}, \texttt{PGB4}, and \texttt{PGB6} systems conducted fine-tuning of the last two, four, and six layers while freezing the other layers, respectively. The \texttt{PGB2T} uses multi-downstream tasks of both TTS and tone prediction. In all systems, only an output from the last layer was used as input feature to TTS. Note that the \texttt{PGB2T} system did not use tone labels during prediction. \texttt{PGB2MC} was to distinguish the contribution of the graphemes by masking all graphemes during fine-tuning and prediction.

We constructed another pre-trained BERT model that uses only phonemes during pre-training, fine-tuning and prediction as a baseline to clarify the contribution of the graphemes. In this system, pre-training, fine-tuning, and prediction scheme were same as PnG~BERT except that all graphemes were masked out. We trained phoneme BERT up to 1.0M steps with 80 batch size with linear decay started from $5.0 \times 10^{-5}$ with the Adam optimizer with L2 regularization. We fine-tuned the last two layers of phoneme BERT with TTS task. We referred to this system as \texttt{PB2MC}.

\begin{itemize}
    \item[7)] \texttt{PB2MC}: Phoneme BERT that uses only phonemes as inputs during pre-training and fine-tuning. The last two layers of BERT encoder are fine-tuned.
\end{itemize}

We prepared five baseline systems with different encoder and input representations.

\begin{itemize}
    \item[8)] \texttt{PGBN}: Non-pretrained PnG~BERT as the encoder with grapheme and phoneme labels
    \item[9)] \texttt{TAC}: Tacotron2's encoder with phoneme labels
    \item[10)] \texttt{TACT}: Tacotron2's encoder with phoneme and tone labels
    \item[11)] \texttt{BTAC2A}: Tacotron2's phoneme encoder and pre-trained character BERT encoder. Its two sources are aligned to target with dual source attention \cite{BLP:journals/corr/ZophK16}.
    \item[12)] \texttt{BTAC2B}: Tacotron2's phoneme encoder and pre-trained character BERT encoder. Its two sources are aligned to target with dual source biattention \cite{DBLP:conf/nips/McCannBXS17}.
\end{itemize}

The \texttt{PGBN} system used the same framework to conduct TTS with PnG~BERT except for skipping pre-training. It used the same BERT architecture, input representation, and TTS procedure as the other five systems using PnG~BERT.

We used Tacotron2 \cite{Shen2017} as the baseline method, different from PnG~BERT. We constructed two Tacotron2 models without pre-training: \texttt{TAC} that uses the phoneme label and \texttt{TACT} that uses phoneme and tone labels as inputs. The phoneme and tones were concatenated after being processed by the pre-net layer to feed to the encoder in the \texttt{TACT} system \cite{Zhang2018}.  We used the same phoneme and tone labels that were used in the fine-tuning of TTS and tone prediction for the PnG~BERT-based systems. The encoder of Tacotron2 consists of three convolutional layers and a bidirectional LSTM layer. Note that \texttt{TAC} is a baseline comparable to the PnG~BERT systems considering that the PnG~BERT systems did not use tone labels during prediction, and \texttt{TACT} worked as the upper bound as it used tone labels during prediction.

We also included two Tacotron baseline systems extended with pre-trained character BERT encoder. These systems combines the character BERT encoder and phoneme encoder by dual source attention \cite{BLP:journals/corr/ZophK16}. This was a common method to combine BERT with Tacotron \cite{DBLP:conf/interspeech/HayashiWTTTL19, Aso2020}, and we referred to the system as \texttt{BTAC2A}. This method did not consider alignment between the two sources, i.e. phoneme features and character features. The \texttt{BTAC2B} system used dual source biattention to learn alignment between the two sources \cite{DBLP:conf/nips/McCannBXS17}. With biattention, the model learned to align the two sources, and then the aligned sources are aligned to targets to predict outputs. Guided attention loss \cite{Tachibana2018} were used for alignments between BERT features and target mel-spectrogram to enforce monotonicity. We used a publicly available pre-trained Japanese BERT model that used characters as input representation and was trained with word-level masking \footnote{BERT-base\_mecab-ipadic-char-4k\_do-whole-word-mask published at https://github.com/cl-tohoku/bert-japanese}. We fine-tuned the last two layers of BERT encoder.

All these systems used the same decoder as that of Tacotron2 \cite{Shen2017}: two-layer LSTM decoder, attention layer with forward attention \cite{Zhang2018}, and two-layer pre-net bottleneck. The forward attention enabled faster training by enforcing a monotonic structure of alignment \cite{Zhang2018}. The decoder pre-net layer prevented the decoder from excessively depending on autoregressive feedback by applying dropout \cite{Wang2017}.

\subsection{Fine-tuning conditions}

\subsubsection{Fine-tuning with TTS task}

The TTS models were fine-tuned with a batch size of 60 and learning rate of $1.0 \times 10^{-6}$ until validation loss stopped improving. The \texttt{PGB0} system was trained up to 488K steps by using pre-trained PnG~BERT parameters. The \texttt{PGB2} system was trained up to 272K steps by warm-starting from parameters of \texttt{PGB0}. The \texttt{PGB4} system was trained up to 90K steps by warm-starting from parameters of \texttt{PGB2}. The \texttt{PGB6} system was trained up to 142K steps by warm-starting from parameters of \texttt{PGB4}. The \texttt{PGBN} system was trained up to 282K steps from scratch.  
The \texttt{PGB2MC} system was fine-tuned up to 72k steps by warm-starting from parameters of \texttt{PGB2} . The \texttt{PB2MC} system was fine-tuned up to 156k steps by freezing all layers and then was fine-tuned  up to 32k steps by unfreezing the last two layers. The  \texttt{BTAC2A} system was fine-tuned up to 80k steps by freezing all layers and then was fine-tuned  up to 20k steps by unfreezing the last two layers. The \texttt{BTAC2B} system was fine-tuned up to 64k steps by freezing all layers and then was fine-tuned  up to 26k steps by unfreezing the last two layers.
We used the same optimizer, i.e., Adam with L2 regularization, to fine-tune a TTS model \cite{DBLP:journals/corr/KingmaB14}.

We extracted an 80-dimensional mel-spectrogram from a waveform with a 24k sampling rate as a target acoustic feature. We used Parallel Wave GAN \cite{DBLP:conf/icassp/YamamotoSK20} to generate a waveform from the mel-spectrogram.

We used JSUT \cite{DBLP:journals/corr/abs-1711-00354} as the speech corpus to fine-tune TTS models. We used phoneme labels transcribed from speech  \footnote{kana\_level3 in https://github.com/sarulab-speech/jsut-label .}. We split the data into train, validation, and test sets with 5,386, 499, and 500 sentences, respectively.

\subsubsection{Fine-tuning with TTS and tone prediction task}
\label{subsec:fine-tuning-with-tone-prediction}

To capture accent information, the PnG~BERT model was fine-tuned by the tone prediction task in the \texttt{PGB2T} system. In this setting, a PnG~BERT model predicts tone labels from the phoneme segment of an input sequence. 

The \texttt{PGB2T} system was trained up to 180K steps by warm-starting from parameters of \texttt{PGB0} with the Adam optimizer with L2 regularization. The fine-tuning of tone prediction was performed at the same time with TTS fine-tuning. We found that the multitask learning of TTS and tone prediction was essential to incorporate tone prediction to PnG~BERT. Separating the fine-tuning by TTS and tone prediction tasks in two stages failed to learn the alignment between input features from PnG~BERT and target speech during TTS training.

We used the same corpus as that of TTS, namely, JSUT corpus \cite{DBLP:journals/corr/abs-1711-00354} to fine-tune with tone prediction \footnote{The accent labels were obtained from https://github.com/sarulab-speech/jsut-label .}. The data split setting for tone prediction was the same as that of TTS. We used the X-JToBI \cite{Maekawa2002} format to represent the tone labels. The tone label contains the following pitch contour patterns: L (neutral low), and H (neutral high), \%L (neutral low at the start of accentual phrase), L\% (neutral low at the end of accentual phrase), and A (accent nucleus position). The tone labels were aligned to phoneme labels.

\subsection{Evaluation Methods}

\subsubsection{Pre-training evaluation}

We evaluated the performance of pre-training with three metrics: accuracies of MLM, G2P, and P2G as in \cite{jia21_interspeech}. The accuracy of MLM was computed on the basis of the MLM objective with random masking, the same as in the pre-training. The accuracy of G2P was computed by masking the entire phoneme segment, and the accuracy of P2G was computed by masking the entire grapheme segment. 

\subsubsection{Objective evaluations}

We measured attention error rate (AER) \cite{YASUDA2021101183}, character error rate (CER), tone prediction accuracy (TA), phrase prediction accuracy (PA), and accentual nucleus prediction accuracy (AA) as objective metrics. 

The AER was used to evaluate how easily the feature from PnG~BERT was aligned to speech. The AER was calculated on the basis of the number of non-monotonic attention distributions that contain discontinuities or overestimation of duration. We counted discontinuities over four encoder time steps and duration over 30 decoder time steps as alignment errors. 

The CER was used to evaluate the phonetic correctness of synthetic speech. The CER refers to the character-level Levenshtein edit distance between reference texts and hypothesis texts transcribed by automatic speech recognition (ASR). We used ESPNet \cite{DBLP:conf/interspeech/WatanabeHKHNUSH18} for ASR \footnote{We used a publicly available ASR model: https://zenodo.org/record/4304245\#.Yac84FORVqs .}. 

The TA, PA, and AA were used to evaluate the abundance of tone-related information in encoded features. The TA is prediction accuracy of all tone labels described in Section \ref{subsec:fine-tuning-with-tone-prediction}. The PA is the prediction accuracy of start and end of accentual phrase (\%L and L\%). The AA is the prediction accuracy of accent nucleus (A). They were measured by conducting linear classification against fixed features from the encoder of PnG~BERT or Tacotron. In the case of the \texttt{BTAC2A} system, concatenation of outputs from phoneme encoder and BERT encoder were used as inputs for the tone classifier. In the case of the \texttt{BTAC2B} system, concatenation of outputs from phoneme encoder and aligned BERT encoder by biattention were used as inputs for the tone classifier.

\subsubsection{Subjective evaluations}

We conducted a listening test to evaluate the synthetic speech \footnote{Our audio sample page can be found here: https://todalab.github.io/yasuda-japanese-pngbert-samples/ .}. We included the ten systems described in Section \ref{subsec:tts-systems} except for \texttt{PGB2MC}, \texttt{PB2MC}, \texttt{BTAC2A}, and \texttt{BTAC2B} in addition to natural samples (\texttt{NAT}) and analysis-by-synthesis (\texttt{ABS}) in the listening test. Here, ABS meant synthetic samples from neural vocoder given ground truth mel-spectrogram. We asked two questions to listeners. The first question was about naturalness in the five-scale mean opinion score (MOS): very bad, bad, acceptable, good, and very good. The second question was about accent correctness in four-scale MOS: totally wrong, somewhat wrong, somewhat correct, and totally correct. Every sample was evaluated five times. We collected 25,000 evaluations in total from 248 Japanese listeners. We checked the statistical significance with the Mann-Whitney rank test \cite{Mann-WhitneyRankTest}.

\subsection{Experimental Results}

\subsubsection{Pre-training results}

\begin{figure}[!t]
    \begin{center}
    {\includegraphics[width=0.8\columnwidth]{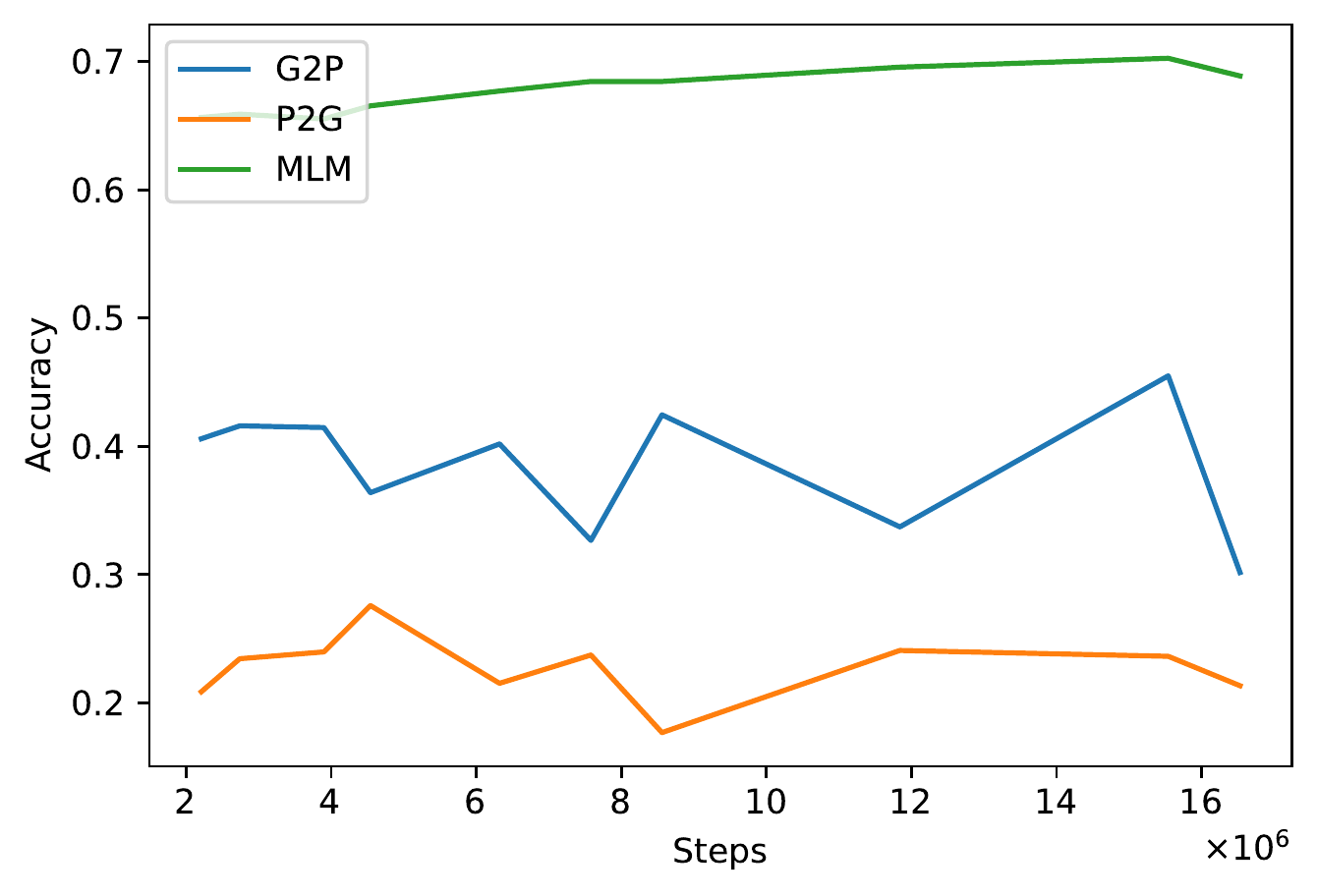}}
    \end{center}
    \caption{History of pre-training accuracy of a PnG~BERT model for validation set. MLM refers to accuracy under random masking as training time. G2P refers to accuracy under masking all phonemes. P2G refers to accuracy under masking all graphemes.}
    \label{fig:pngbert-accuracy}
\end{figure}

Figure \ref{fig:pngbert-accuracy} shows the accuracies of MLM, G2P, and P2G in the validation set as the performance of pre-training. The accuracy of MLM consistently improved during pre-training up to 70.3\% until overfit. On the other hand, the accuracies of G2P and P2G fluctuated during pre-training, which were 45.5\% and 23.6\% at the model with the highest MLM accuracy, respectively. It seemed that the performances of G2P and P2G had a trade-off relationship: when one metric increased, the other decreased. The magnitudes of these values were somewhat consistent with \cite{jia21_interspeech}. The relatively low performances of G2P and P2G indicated that the features captured by PnG~BERT were not dominant in surface-form information with our masking strategy of pre-training, which was expected to contain syntactic and semantic information \cite{jia21_interspeech}. 

\subsubsection{TTS results}

Table \ref{tbl:tts-error-rates} shows AER, CER, TA, PA, and AA from the TTS systems. The AER and CER showed roughly the same trends: the systems that had high AER also showed high CER. When the number of fine-tuning layers in PnG~BERT was increased, both AER and CER improved, which indicated that features from PnG~BERT contained information more aligned to speech. The high AER and CER from the \texttt{PGB0} system showed that the pre-trained PnG~BERT without fine-tuning provides abundant textual features and less spoken features, and it was consistent with the relatively low G2P accuracy in the pre-training. The \texttt{PGBN} system showed high AER and CER although it fine-tuned all layers in PnG~BERT. This implied that pre-training was essential for PnG~BERT to learn stable alignment between phonemes and speech for TTS.

As for tone prediction accuracy, pre-trained PnG~BERT systems (\texttt{PGB0}-\texttt{6}) showed high accuracies. They showed high PA but low AA, which indicated that the high TA values mainly came from rich accentual phrase information in the encoded features, and accentual nucleus information was not sufficiently captured in the features. The system with tone prediction (\texttt{PGB2T}) had high AA, which indicating that its feature contained both accentual phrase and nucleus information. 
The masking graphemes (\texttt{PGBMC}) resulted in degradation of TA from 73.1\% to 63.3\%. The degradation was mainly caused by significant decrease of PA from 73.5\% to 38.4\%, suggesting that phrase information depends largely on contextual information of graphemes rather than phonemes. 
The low TA and PA of the PnG~BERT system without pre-training (\texttt{PGBN}) indicated that pre-training was essential to capture information related to accentual phrase. Tacotron, which did not perform pre-training, also failed to capture sufficient information related to accentual phrase, as indicated by its low TA and PA. The high tone prediction accuracy of \texttt{TACT} was because tone labels were given as input. 
Both PnG~BERT systems that use only phonemes (\texttt{PGB2MC} and \texttt{PB2MC}) showed low TA compared to the system using both characters and phonemes (\texttt{PGB2}). The performance drop of TA was mainly caused by degradation of PA. This indicated that contextual information encoded in character part enriched phrase-related information. TA from these systems were similar values to that of the Tacotron baselines. This suggested that pre-training or fine-tuning with only phonemes did not help to encode tone-related information.
Both Tacotron systems extended with character BERT (\texttt{BTAC2A} and \texttt{BTAC2B}) showed similar values of TA, which were lower compared to TA from PnG~BERT systems. We thought that the no improvement of TA from BERT incorporated Tacotron systems were caused by poor alignments between phoneme features and BERT features. We found that it was hard to learn alignment between BERT features and mel-spectrogram in both \texttt{BTAC2A} and \texttt{BTAC2B} systems, and alignment between BERT features and phoneme features in the \texttt{BTAC2B} system even though various techniques such as forward attention, guided attention, and biattention were used to enforce robust alignments. The attention distributions about BERT features were mostly blurred or nonmonotonic. These results were consistent with the existing works \cite{DBLP:conf/interspeech/HayashiWTTTL19, Aso2020}. We thought learning alignment proprly between phoneme features and BERT features was important to predict tones from the contextual features because tone labels were designed in phoneme-level, and high TA from PnG~BERT systems came from learning accurate alignments between phonemes and characters by utilizing large-scale text resources during pre-training.

\begin{table}[t]
\setlength{\tabcolsep}{4pt}
\caption{Alignment error rate (AER), character error rate (CER), tone prediction accuracy (TA), phrase prediction accuracy (PA), and accentual nucleus prediction accuracy (AA) of TTS systems. Pred. denotes prediction.}
\label{tbl:tts-error-rates}
\begin{center}
\begin{tabular}{|ccccrrrrr|}\hline
\multirow{2}{*}{System} & Pre- & Fine- & \multirow{2}{*}{Tone} & \multirow{2}{*}{AER$\downarrow$} & \multirow{2}{*}{CER$\downarrow$} & \multirow{2}{*}{TA$\uparrow$} & \multirow{2}{*}{PA$\uparrow$} & \multirow{2}{*}{AA$\uparrow$} \\
~ & training & tuning & ~ & ~ & ~ & ~ & ~ & ~\\\hline
\texttt{PGB0} & \checkmark & - & - & 13.6 & 29.4 & 72.8 & 74.1 & 20.6\\
\texttt{PGB2} & \checkmark & 2 layers & - & 5.2 & 19.7 & 73.1 & 73.5 & 22.6\\
\texttt{PGB4} & \checkmark & 4 layers & - & 4.2 & 19.0 & 72.9 & 73.0 & 22.4\\
\texttt{PGB6} & \checkmark & 6 layers & - & 3.6 & 18.6 & 70.0 & 66.8 & 16.2\\
\texttt{PGBN} & - & All & - & 13.2 & 44.2 & 58.4 & 27.8 & 26.0\\
\texttt{PGB2T} & \checkmark & 2 layers & Pred. & 7.6 & 23.0 & 81.6 & 80.5 & 56.7\\
\texttt{PGB2MC} & \checkmark & 2 layers & - & 6.8 & 25.5 & 66.5 & 56.5 & 23.4\\
\texttt{PB2MC} & \checkmark & 2 layers & - & 5.6 & 28.8 & 64.1 & 38.3 & 27.0\\
\texttt{TAC} & - & All & - & 4.4 & 21.0 & 66.0 & 56.2 & 20.9\\
\texttt{TACT} & - & All & Label & 5.4 & 19.9 & 88.5 & 82.7 & 69.0\\
\texttt{BTAC2A} & \checkmark & 2 layers & - & 3.0 & 21.8 & 65.1 & 49.9 & 11.4\\
\texttt{BTAC2B} & \checkmark & 2 layers & - & 20.6 & 22.0 & 60.7 & 48.2 & 17.1\\
\texttt{ABS} & ~ & ~ & ~ & - & 21.0 & - & - & -\\
\texttt{NAT} & ~ & ~ & ~ & - & 24.9 & - & - & -\\\hline
\end{tabular}
\end{center}
\end{table}

\begin{figure}[!t]
    \begin{center}
    \begin{subfigure}[t]{\columnwidth}
    {\includegraphics[width=\columnwidth]{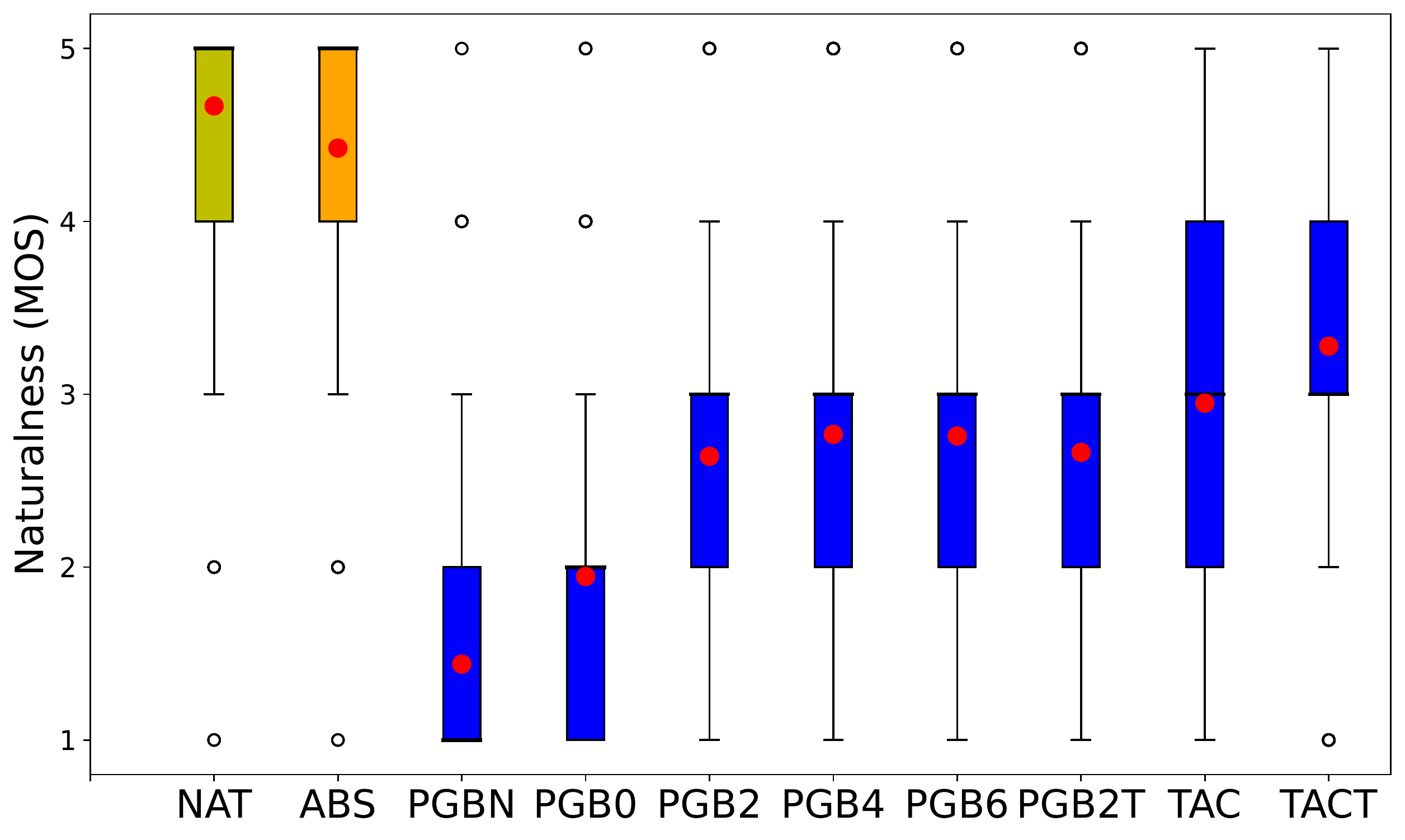}}
    \subcaption{Naturalness}
    \end{subfigure}
    \vspace{2mm}

    \begin{subfigure}[t]{\columnwidth}
    {\includegraphics[width=\columnwidth]{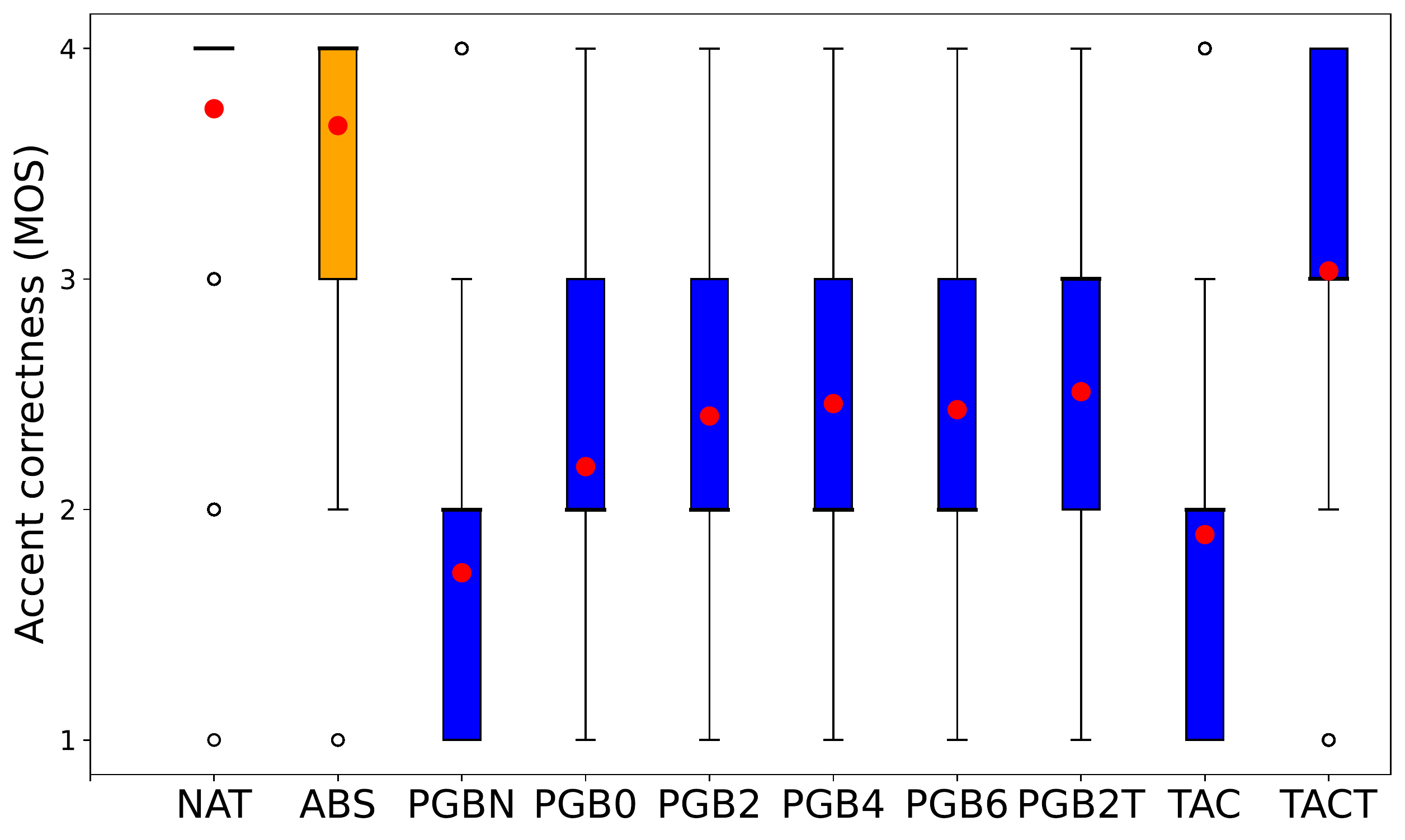}}
    \subcaption{Accent correctness}
    \end{subfigure}
    \end{center}
    \caption{Results of listening test.}
    \label{fig:listening-test}
\end{figure}

Figure \ref{fig:listening-test} shows results of the listening test on (a) naturalness and (b) accent correctness. It showed that pre-training was crucial, considering the significant improvement of naturalness from $1.44\pm0.02$ (\texttt{PGBN}) to $1.95\pm0.03$ (\texttt{PGB0}). Fine-tuning was also essential, considering the significant improvement of naturalness from $1.95\pm0.03$ (\texttt{PGB0}) to $2.64\pm0.03$ (\texttt{PGB2}). Increasing the number of fine-tuned layers slightly improved naturalness from $2.64\pm0.03$ (\texttt{PGB2}) to $2.77\pm0.03$ (\texttt{PGB4}), but there was no significant difference between the four and six layers, indicating that four layers were sufficient for fine-tune. The tone prediction did not help in improving naturalness, because there was no significant difference between \texttt{PGB2} and \texttt{PGB2T}. The PnG~BERT-based systems did not match the naturalness of the Tacotron that had MOS of $2.95\pm0.03$ (\texttt{TAC}). We found that samples from the PnG~BERT-based systems had lower fidelity than samples from the Tacotron systems. We consider that the low fidelity was caused by the difficulty of optimizing the fine-tuning of PnG~BERT, as indicated by long fine-tuning time requirements and relatively high loss values. The Tacotron with accent labels (\texttt{TACT}) had the highest MOS score of $3.28\pm0.03$ among TTS systems as expected of upper bound system.

For accent correctness, pre-training and fine-tuning were crucial as well. There was no significant difference among systems with different numbers of fine-tuned layers. The tone prediction slightly improved accent correctness from $2.41\pm0.03$ (\texttt{PGB2}) to $2.51\pm0.03$ (\texttt{PGB2T}). The pre-trained PnG~BERT based systems significantly outperformed Tacotron, which had MOS of $1.89\pm0.03$ (\texttt{TAC}), in terms of accent correctness in contrast to the results on naturalness. This suggests that pre-trained features from PnG~BERT were helpful in inferring pitch accents. The PnG~BERT without pre-training, which showed a low score of $1.73\pm0.03$ (\texttt{PGBN}), contained graphemes in inputs, so what helped in inferring pitch accents was not surface form information but presumably syntactic and semantic information. Still, accents inferred by the pre-trained PnG~BERT-based systems were not sufficiently accurate compared to accent labels, as indicated by their lower scores than \texttt{TACT}, which used accent labels during prediction. 
The Tacotron with accent labels (\texttt{TACT}) showed relatively high MOS score of $3.04\pm0.03$,
which indicated that Japanese listeners were sensitive to pitch accent.
We consider that the word coverage of training speech and tone label data used for fine-tuning limited the correctness of inferred accents, because the only phase in which accent nucleus positions can be learned by PnG~BERT was fine-tuning in our method. Considering that accent nucleus positions are the spoken features of words, we think that pre-training with tone labels or fine-tuning with large-scale speech data would be required to improve the accent correctness of Japanese PnG~BERT.

\section{Related works}
\label{sec:related-works}

An early attempt to utilize an unsupervised textual representation for TTS is the vector space model, which is an information retrieval method to derive linguistic feature vectors from texts by singular value decomposition \cite{watts2013unsupervised}. The objective of the vector space model is to improve the language versatility of TTS by replacing text front-end, which is traditionally dependent on many language-specific features. However, the vector space model is not helpful for DNN-based TTS \cite{lu13_ssw}.

Neural-network-based pre-training methods replace the vector space model by using much larger text resources.
The neural-network-based pre-training is useful for TTS in ideographic languages to overcome data sparsity caused by character diversity. For example, pre-trained character embedding is used for a front-end model in Maindarin TTS \cite{DBLP:conf/icassp/PanYZLZMW20}. Pre-trained linguistic encoder is used to enable to take ideographic characters directly as input in Mandarin TTS by introducing G2P objective \cite{Li2019KnowledgeBased}.

Word embedding \cite{DBLP:conf/interspeech/MikolovKBCK10, DBLP:journals/corr/abs-1301-3781} and contextual word embedding such as BERT \cite{DBLP:conf/naacl/DevlinCLT19} are  widely used unsupervised word representation learning methods for general purposes. 
Word embedding is mainly used to improve prosody of synthetic speech in TTS \cite{DBLP:conf/icassp/WangQSHZ15, DBLP:journals/corr/abs-1906-07307, DBLP:journals/corr/abs-1901-00707, DBLP:conf/interspeech/HayashiWTTTL19, DBLP:conf/icassp/XiaoHMS20, DBLP:conf/icassp/XuSZZHZ21, nakata21_ssw}. Moreover, the word embedding is suitable for prosody modeling in variational autoencoder (VAE)-based TTS \cite{DBLP:conf/interspeech/TyagiNRDL20, DBLP:conf/interspeech/KenterSC20, DBLP:conf/icassp/KarlapatiAHMJKD21}. Word embedding is also used to improve data efficiency under small amounts of training speech data \cite{DBLP:conf/icassp/ChungWHZS19}.

BERT can not only be used to extract fixed contextual word embedding features, but can also be customized to fit the downstream task. There are three strategies to customize a BERT model: (1) fine-tuning, (2) input representation, and (3) masking strategy. Fine-tuning enables the fitting of features to a downstream task by tuning part of the parameters of the pre-trained model. It is known that fine-tuning improves the performance of BERT in downstream tasks similar to pre-training methods in the NLP domain \cite{DBLP:conf/rep4nlp/PetersRS19}. Recent TTS works using BERT perform fine-tuning \cite{DBLP:journals/corr/abs-1906-07307, DBLP:conf/interspeech/KenterSC20, DBLP:conf/icassp/KarlapatiAHMJKD21, nakata21_ssw, jia21_interspeech}. Some studies show that the fine-tuning of BERT with TTS improves the naturalness of synthetic speech \cite{DBLP:conf/interspeech/KenterSC20, nakata21_ssw, jia21_interspeech}. In this study, we fine-tuned PnG~BERT with TTS and tone prediction and confirmed the effectiveness of fine-tuning.

The design of the input representation of BERT for TTS includes words, subwords, characters, phonemes, and speech. Pre-training in the word domain obtains semantic information, and pre-training in the character domain obtains surface information, and pre-training in the subword domain is considered between them. Words \cite{DBLP:conf/icassp/XuSZZHZ21, DBLP:conf/icassp/XuSZZHZ21} and subwords \cite{DBLP:journals/corr/abs-1906-07307, DBLP:conf/interspeech/TyagiNRDL20, DBLP:conf/interspeech/KenterSC20, DBLP:conf/icassp/KarlapatiAHMJKD21, nakata21_ssw} are commonly used representations for BERT. In \cite{DBLP:conf/icassp/XiaoHMS20}, characters were used as input to BERT for Mandarin TTS. PnG~BERT \cite{jia21_interspeech} combines both characters and phonemes to pre-train BERT to capture both surface and phoneme information to apply to TTS. This idea to combine subwords and phonemes is also investigated in spoken language understanding \cite{chen21g_interspeech}. In \cite{DBLP:conf/icassp/ChenDWSH21}, mel-spectrogram was used to pre-train BERT, which is used as a pre-net network \cite{Wang2017} in TTS. 

The masking strategy can be designed specifically for a downstream task. PnG~BERT \cite{jia21_interspeech} applies word-level masking on both character and phoneme pairs. \cite{chen21g_interspeech} uses ``one-mode masking'' was used on subwords and phonemes, which is to mask the entire subword or phoneme segment equivalent to the G2P or P2G task.

\section{Conclusion}
\label{sec:conclusion}

In this work, we investigated the effects of features captured by PnG~BERT on Japanese TTS by modifying the fine-tuning condition to determine the conditions helpful in rendering pitch accents. We manipulated the content of PnG~BERT features from being text-oriented to speech-oriented by changing the number of fine-tuned layers during TTS. In addition, we taught PnG~BERT pitch accent information by fine-tuning with tone prediction as an additional downstream task. Our experiment showed that pre-training and fine-tuning were essential for both naturalness and accent correctness for Japanese TTS. Increasing the number of fine-tuned layers improved alignment and character error rates, but did not contribute considerably to naturalness. PnG~BERT provided better accent correctness than the Tacotron baseline, although all the PnG~BERT-based systems were inferior to Tacotron in terms of naturalness because of their low fidelity. Fine-tuning with tone prediction considerably improved tone prediction accuracies from captured features, but its improvements were limited in the subjective evaluation of accent correctness. Overall, the features of PnG~BERT captured by pre-training contained information helpful in inferring pitch accent, and fine-tuning by TTS enriched pitch accent information for the PnG~BERT features.

Our future works include the improvement of the low fidelity of generated speech and long fine-tuning time of PnG~BERT. It would be interesting to find the optimal masking strategy to obtain higher fidelity and fine-tuning efficiency of Japanese PnG~BERT by changing the masking strategy to control the balance of surface form, syntactic, and semantic information. We will also work on the further improvement of Japanese PnG~BERT to infer pitch accents. One interesting way to improve the capture of pitch accent features is pre-training PnG~BERT with accent labels instead of fine-tuning by tone prediction. Another interesting way is to fine-tune PnG~BERT with large-scale speech data.

\section*{Acknowledgments}

This study was partly supported by a project, JPNP20006, commissioned by NEDO.

\bibliographystyle{./IEEEtran.bst}
\bibliography{reference.bib}


\section{Biography Section}

\begin{IEEEbiographynophoto}{Yusuke Yasuda} received his B.S. and M.S. degrees at Waseda University, Japan, in 2012 and 2014, respectively. He received his Ph.D degree from the graduate university for advanced studies, SOKENDAI, Japan, in 2021. He was a research assistant at the National Institute of Informatics, Japan, from 2018 to 2021. Since 2021, he has been an assistant project professor at the Information Technology Center, Nagoya University.
He received the 20th Best Student Paper Award from the Acoustical Society of Japan.
His research interests include statistical machine learning and speech synthesis.
\end{IEEEbiographynophoto}

\begin{IEEEbiographynophoto}{Tomoki Toda} received his B.E. degree from Nagoya University, Japan, in 1999 and his M.E. and D.E. degrees from Nara Institute of Science and Technology (NAIST), Japan, in 2001 and 2003, respectively. He was a Research Fellow of the Japan Society for the Promotion of Science, from 2003 to 2005. Then, he was an assistant professor from 2005 to 2011 and an associate professor from 2011 to 2015 at NAIST. Since 2015, he has been a professor at the Information Technology Center, Nagoya University. He received more than 15 article/achievement awards, including the IEEE SPS 2009 Young Author Best Paper Award and the 2013 EURASIP-ISCA Best Paper Award (Speech Communication Journal). His research interests include statistical approaches to sound media information processing.
\end{IEEEbiographynophoto}

\vfill

\end{document}